\documentclass[12pt]{article}
\usepackage{epsfig}

\usepackage{amssymb}
\usepackage{amsmath}
\usepackage{amsfonts}

\usepackage{amsthm}

\theoremstyle{theorem}

\newtheorem{thm}{Theorem}

\theoremstyle{definition}
\newtheorem{de}{Definition}
\newtheorem{pozn}{Remark}

\def\bp{\begin{proof}}
\def\ep{\end{proof}}

\def\be{\begin{equation}}
\def\ee{\end{equation}}
\def\ba{\begin{array}{c}}
\def\ea{\end{array}}

\newcommand{\bea}{\begin{eqnarray}}
\newcommand{\eea}{\end{eqnarray}}

\newcommand{\bbr}{\br\!\br}

\newcommand{\kt}{\rangle}
\newcommand{\br}{\langle}
\begin{document}

\titlepage

\vspace{.35cm}

 \begin{center}{\Large \bf

Symbolic-manipulation constructions of Hilbert-space metrics in
quantum mechanics

  }\end{center}

\vspace{10mm}

 \begin{center}

 {\bf Miloslav Znojil}

 \vspace{3mm}
Nuclear Physics Institute ASCR,

250 68 \v{R}e\v{z}, Czech Republic

{e-mail: znojil@ujf.cas.cz}

%\vspace{3mm}

% http://www14.in.tum.de/CASC2011/

%\today, znojilcasc.tex

\end{center}

\vspace{5mm}

\section*{Abstract\footnote{Work supported by the GA\v{C}R grant Nr.
P203/11/1433, by the M\v{S}MT ``Doppler Institute" project Nr.
LC06002 and by the Inst. Res. Plan AV0Z10480505. To be presented to
the Computer Algebra in Scientific Computing (CASC 2011) conference
in Kassel, Germany, during September 5-9, 2011.}}

The problem of the determination of the Hilbert-space metric
$\Theta$ which renders a given Hamiltonian $H$ self-adjoint is
addressed from the point of view of applicability of
computer-assisted algebraic manipulations. An exactly solvable
example of the so called Gegenbauerian quantum-lattice oscillator is
recalled for the purpose. Both the construction of suitable
$\Theta=\Theta(H)$ (basically, the solution of the Dieudonne's
operator equation) and the determination of its domain of positivity
are shown facilitated by the symbolic algebraic manipulations and by
MAPLE-supported numerics and graphics.

\section{Introduction}

In the series of papers \cite{fund} - \cite{posledni} we interpreted
the Dieudonn\'e's quasi-Hermiticity relation \cite{Dieudonne}
 \be
 H^\dagger\,\Theta=\Theta\,H\,
  \label{dieudo}
 \ee
as an operator equation which connects a given, ``input" quantum
Hamiltonian $H$ with an unknown, ``output" operator $\Theta$ called
the Hilbert-space metric of the quantum system in question
(\cite{Geyer}; cf. also a more detailed explanation of the
underlying physics in Appendix A below).

In all of our papers we felt addressed by the underlying physics
(i.e., by quantum mechanics in its form described, say, in Refs.
\cite{Carl,SIGMA}) and did not pay too much attention to the
underlying constructive mathematics. In our present paper we intend
to fill this gap by redirecting our attention to the
computer-assisted symbolic-manipulation background of our results.

We shall assume that both of the operators $H$ and $\Theta$ in
Eq.~(\ref{dieudo}) are defined in a vector space ${\cal V}$ with the
Dirac-ket elements $|\psi\kt \in {\cal V}$. For the sake of
simplicity we shall further assume that ${\rm dim}\,{\cal
V}=N<\infty$. In such a setting one may select various toy-model
matrices $H^{(N)}$ and construct the eligible metrics $\Theta$. For
such a purpose we shall choose the Gegenbauerian quantum $N-$site
lattices and study $H=H^{(N)}(a)$ and $\Theta=\Theta^{(N)}(a)$ of
Ref.~\cite{gegenb} (cf. section \ref{gegia}). The reasons of a
facilitated algebraic tractability of Eq.~(\ref{dieudo}) will be
then clarified in section \ref{refoa}, with some complementary
numerical aspects mentioned in section \ref{duik}.

\section{Gegenbauerian quantum $N-$site lattices\label{gegia}}

The general Hermitian conjugation prescribed by Eq.~(\ref{nediraco})
of Appendix A must be compatible with the principles of Quantum
Mechanics. This means that, formally speaking \cite{SIGMA}, our
choice of the metric $\Theta$ must guarantee  the Hermiticity of the
observables (i.e., in our present paper, just of the Hamiltonian
$H$) with respect to this conjugation \cite{Geyer}. Such a
requirement implies the necessity of the validity of the
above-mentioned relation (\ref{dieudo}).

The latter relation will be called here, for the sake of brevity,
Dieudonn\'e equation. As long as this is the matrix equation, it
seems to be an overdeterminate constraint. Its $N^2$ items have to
be satisfied by the mere $N(N+1)/2$ independent matrix elements of
the general $N$ by $N$ real and symmetric matrix $\Theta$ with
positive eigenvalues. Via a deeper study of an illustrative example
taken from Ref.~\cite{gegenb} we intend to demonstrate that the
situation is much more user friendly.

First of all, the Hermitian conjugation may be perceived as a
symmetry of Eq.~(\ref{dieudo}) so that just its upper-triangular
nontrivial subset remains relevant. This implies that the whole set
of equations is in fact underdeterminate. {\em A priori}, the
independent solutions $\Theta$ will  form an $N-$parametric family.
This observation is compatible with the explicit constructive
results published in Ref.~\cite{gegenb}.

Secondly, the message delivered by Ref.~\cite{gegenb} was aimed at
the physics audience. Our present study will complement these
results by their more systematic derivation and by the more explicit
explanation of their formal structure. Keeping this aim in mind we
shall consider the $N-$dimensional matrix Schr\"{o}dinger equation
 \be
 H^{(N)}\,|\psi_n^{(N)}\kt=E_n^{(N)}\,|\psi_n^{(N)}\kt\,
 \label{SEgeg}
 \ee
with the prescribed bound-state eigenvectors
 \be
 |\psi_n^{(N)}\kt=\left (
  \ba
 \br 0|\psi_n^{(N)}\kt=G(0,a,E_n)\\
 \br 1|\psi_n^{(N)}\kt=G(1,a,E_n)\\
 \vdots\\
 \br N-1|\psi_n^{(N)}\kt=G(N-1,a,E_n)
 \ea
 \right )\,
 \label{input}
 \ee
where, in the notation of MAPLE \cite{Maple}, the symbol $G(n,a,x)$
denotes the $n-$th Gegenbauer polynomial $G(n,a,x)$ equal to
polynomial $C^a_n(x)$ in the notation of Ref.~\cite{Ryzhik} or to
$C^{(a)}_n(x)$ according to Ref.~\cite{Stegun}. Under such an
assumption, naturally, the explicit form of the related Hamiltonian
is the tridiagonal array
 \be
 \left[ \begin {array}{cccccc}
 0&1/2\,{a}^{-1}&0&0&\ldots&0
\\
\noalign{\medskip}2\,{\frac {a}{2\,a+2}}&0& \left( 2\,a+2 \right) ^{
-1}&0&\ldots&\vdots
\\
\noalign{\medskip}0&{\frac {2\,a+1}{2\,a+4}}&0& \left(
2 \,a+4 \right) ^{-1}&\ddots&0
\\
\noalign{\medskip}0&0&{\frac
{2\,a+2}{2\,a+ 6}}&\ddots& \ddots&0
\\
\noalign{\medskip}\vdots&\ddots&\ddots&\ddots&0&
\left( 2\,a+2N-4 \right) ^{-1}
\\
\noalign{\medskip}0&\ldots&0&0&{\frac {2\,a+N-1}{2\,a+2N-2}}&0
\end {array} \right]
 \label{hamil}
 \ee
which defines the manifestly asymmetric  $N$ by $N$ matrix
$H^{(N)}$.

The validity of such an assignment is equivalent to the standard
three-term recurrences for the Gegenbauer polynomials while the $N$
by $N$ matrix truncation is equivalent to the implicit-equation
identification of the real and non-degenerate spectrum
$\sigma(H^{(N)}) \equiv \{E_n\}$ of the bound-state energies with
the roots of the $N-$th Gegenbauer polynomial,
 \be
 G(N,a,E_n)=0\,.
 \label{secu}
 \ee
Naturally, such a secular equation may be considered solvable with
an arbitrary numerical precision. The only nontrivial task
represented by the complete description of the model will lie in the
choice of a metric $\Theta$ compatible with  Dieudonn\'e
Eq.~(\ref{dieudo}). The method has not been described in
Ref.~\cite{gegenb} because it just consisted in the brute-force
insertion of a general real and symmetric ansatz for $\Theta^{(N)}$
and in the subsequent trial and error analysis of Eq.~(\ref{dieudo})
after its insertion.

In the metric-construction problem the key
difficulties are twofold. Firstly, the ansatz for $\Theta^{(N)}$
contains too many (i.e., $N(N+1)/2$) unknowns and there are no
criteria for the clarification which ones of them should be selected
as the ``optimal" independent set. Even at the very small
integers $N$, preliminary MAPLE-based
brute-force algebraic symbolic-manipulation-solution experiments
starting from Eq.~(\ref{dieudo}) and from a few randomly selected
$N-$plets of the tentative independent matrix elements of $\Theta$
generated just the obscure many-page results for all of the
$N-$ and $a-$dependent matrix elements of $\Theta(H)$.
The second difficulty emerged with the necessity of the
parameter-range-specifying guarantee of the obligatorily
positive-definite nature of any resulting $N$ by $N$ matrix
candidate $\Theta_\alpha(H)$ for the metric (characterized or
distinguished, in general, by a suitable multiindex $\alpha$).

The core of the success (i.e., of the resolution to both of these
parallel algebro-numerical difficulties) has been revealed to lie in
an interactive and iterative approach to both of the problems. In
more detail this approach is to be described in what follows.

\section{The Dieudonn\'e's equation \label{refoa}}

In the light of the old Dieudonn\'e's idea \cite{Dieudonne} it seems
interesting to replace the current textbook Hermiticity property
$H=H^\dagger$ of the current selfadjoint Hamiltonians in quantum
mechanics by the weaker assumption containing a nontrivial ``metric"
$\Theta \neq I$ \cite{Geyer}. In this context relation
(\ref{dieudo}) guarantees the reality of the energies provided only
that we require that the operator $\Theta =\Theta^\dagger$ is,
roughly speaking \cite{Williams}, positive and invertible, i.e.,
tractable as a metric in the Hilbert space of states ${\cal
H}^{(S)}$ where the superscript means ``standard".

The recent growth of popularity and applicability of the quantum
models requiring the unusual metrics $\Theta =\Theta^{(S)}> I$ may
be found reviewed, e.g., in Refs.~\cite{Carl,Dorey,ali}. In our
present paper we circumvent a number of technicalities by studying
just the quantum models defined in finite-dimensional Hilbert
spaces. Thus, we may identify ${\cal H}^{(F)}
\,\equiv\,\mathbb{C}^N$. Moreover, for the sake of definiteness, we
shall only pay attention to the models where the Hamiltonian
matrices possess the tridiagonal real-matrix form.

\subsection{An interactive algebraic-solution technique}

As an illustrative example we shall use the Gegenbauerian quantum
lattice model of Ref.~\cite{gegenb} described in the preceding
section. In fact, in Ref.~\cite{gegenb} we  described the results
showing the feasibility of the brute-force construction of the
complete family of the metrics $\Theta(H)$ admitted by the
Dieudonn\'e's linear algebraic constraints (\ref{dieudo}). In our
present continuation of this effort we intend to provide a deeper
insight in the problem explaining the reasons why our construction
of the metrics appeared to be so successful.

We have to admit that with our very specific choice of the
Gegenbauerian model in Ref.~\cite{gegenb} we were unexpectedly
fortunate. This fact may be demonstrated, say, by the recollection
of the similar constructive attempts based on a different choice of
the $N$ by $N$ ``input" Hamiltonian as reported in
Ref.~\cite{determ}. After the construction of $\Theta(H)$ at the
dimension as low as $N=4$ it has been argued there that the
construction at the very next $N=5$ appeared almost prohibitively
complicated. This is really in contrast with the results of
Ref.~\cite{gegenb} which proved {\em valid at any integer} $N$.

%\subsubsection{The choice of $N=4$}

The core of the dimension-independent universality of the
above-mentioned Gegenbauerian result may be seen in the combination
of the extremely simple bidiagonal form of the Hamiltonians
$H^{(N)}(a)$ with the comparably simple $a-$dependence of its matrix
elements. The relevance of both of these  ingredients becomes
obvious when we recall the explicit $N=4$ sample of the Hamiltonian
 $$
 H^{(4)}(a)=\left[ \begin {array}{cccc} 0&(2\,a)^{-1}&0&0\\\noalign{\medskip}
\,{\frac {2\,a}{2\,a+2}}&0& \left( 2\,a+2 \right) ^{-1}&0
\\\noalign{\medskip}0&{\frac {2\,a+1}{2\,a+4}}&0& \left( 2\,a+4
 \right) ^{-1}\\\noalign{\medskip}0&0&{\frac {2\,a+2}{2\,a+6}}&0
\end {array} \right]
 $$
together with the general ansatz for the metric
 $$
  \Theta^{(4)}(a)=\left[ \begin {array}{cccc} {\it {k}}&b&c&d\\\noalign{\medskip}b&f&g&h
\\\noalign{\medskip}c&g&m&n\\\noalign{\medskip}d&h&n&j\end {array}
 \right]\,.
 $$
In such a setting we may study the 16-plet of the resulting
relations, out of which Nr. 1, Nr. 6, Nr. 11 and Nr. 16 (i.e.,
diagonal items) remain trivial while the off-diagonal items form an
antisymmetric matrix. Out of the remaining six independent items
(say, Nr. 2, 3, 4, 7, 8, and 12) there is just one (viz., Nr. 4)
which involves just two unknown quantities (viz., $h$ and $c$). This
leads to the decision of taking $d$, $c$, $b$ and ${k}$ as
independent parameters and of eliminating, in the first step, $h$
via item Nr. 4,
$$
h= {\frac {c \left( a+1 \right) }{2\, \left( a+2 \right) a}}\,.
$$

The inspection of the new set of items reveals that the simplest one
is now just Nr. 3 which defines $g$ as a function of $b$ and $d$,
with the next-step Nr. 8 defining $n$ as a function of $d$ and
(newly known) $g$. We are left with the three items Nr. 2, 7 and 12
which couple $({k},f)$, $(f,m)$ and $(m,j)$, respectively. As long
as we decided to use ${k}$ as the fourth unconstrained parameter
this means that in the same order we now define, step by step, the
missing items $f$, $m$ and, ultimately, $j$. The result is complete
yielding
$$
h= 1/2\,{\frac {c \left( a+1 \right) }{ \left( a+2 \right) a}}
$$
$$
g=1/2\,{\frac {ba+3\,b+2\,d{a}^{2}+4\,da+2\,d}{ \left( a+3 \right)
a}}
$$
$$
n=1/2\,{\frac {-6\,da-10\,d+ba+3\,b}{ \left( a+3 \right) a \left(
2\,a+1
 \right) }}
$$
$$
f=1/2\,{\frac { \left( 2\,c{a}^{2}+{\it {k}}\,a+ca+2\,{\it {k}}
\right)
 \left( a+1 \right) }{ \left( a+2 \right) {a}^{2}}}
$$
$$
m=1/2\,{\frac {2\,c{a}^{3}+c{a}^{2}-7\,ca+{\it {k}}\,{a}^{2}+5\,{\it
{k}} \,a+6\,{\it {k}}}{ \left( a+3 \right) {a}^{2} \left( 2\,a+1
\right) }}
$$
$$
j=-1/4\,{\frac {6\,c{a}^{2}+10\,ca-{\it {k}}\,{a}^{2}-5\,{\it
{k}}\,a-6\,{ \it {k}}}{{a}^{2} \left( 2\,a+1 \right)  \left( a+2
\right)  \left( a+1
 \right) }}\,.
$$

\subsection{The case of general $N$}

After the above-explained heuristic exercise we are prepared to
consider the real and symmetric general ansatz for the metric
 \be
  \Theta^{(N)}(a) =\left[ \begin {array}{cccc}
  %\hline
   \theta_{11}&\theta_{12}&\ldots&\theta_{1,N}
  \\
  %\hline
  \theta_{12}
&\theta_{22}&\ddots&\vdots
 \\
 \vdots&\ddots&\ddots&\theta_{N-1,N}
 \\
 \theta_{1,N}&
 \ldots&
\theta_{N-1,N}&\theta_{NN}
 %\\
% %\hline
% 0&\ddots&
% &\ddots&\theta_{N-2,N-2}&\theta_{N-2,N-1}&\vdots
% %\\
%  \vdots&\ddots&\theta_{N-1,N-k-1}
%  &\ldots&\theta_{N-1,N-2}&\theta_{N-1,N-1}&\theta_{N-1,N}
 %\\ 0&\ldots&0&\theta_{N,N-k}&\ldots&\theta_{N,N-1}&\theta_{NN}
% %\\
% %\hline
 \end {array} \right]\,.
 \label{kit}
 \ee
and prove the general result.

\begin{de}
At any $N \geq 2$ the insertion of the $N$ by $N$ Hamiltonian
$H=H^{(N)}(a)$ given by Eq.~(\ref{hamil}) and of the general real
and symmetric matrix ansatz (\ref{kit}) for the metric
$\Theta=\Theta^{(N)}(a)$ defines the $N$ by $N$ matrix
array~(\ref{dieudo}) of the linear Dieudonn\'e equations ${\cal
M}_{i,j}=0$. Its {\bf ordered version} has the form $r_\alpha=0$
with
 \begin{eqnarray}
 %\flushleft
  r_1={\cal M}_{1,N}\,,\nonumber \\
 r_2={\cal M}_{1,N-1}\,,r_3={\cal M}_{2,N}\,,\nonumber \\
 r_4={\cal M}_{1,N-2}\,,r_5={\cal M}_{2,N-1}\,,r_6={\cal M}_{3,N}\,,\nonumber \\
 \ldots\,,\nonumber \\
 r_{(N-1)(N-2))/2+1}={\cal M}_{1,2}\,,\ldots\,, r_{N(N-1)/2}={\cal
 M}_{N-1,N}\,.
 \label{tiral}
  \end{eqnarray}

\end{de}

\begin{thm}
In terms of the freely variable $N-$plet of the real initial
parameters $ \Theta_{1j}$, $j = 1,2,\ldots,N$ the Dieudonn\'e
equation in its ordered version (\ref{tiral}) defines, step by step,
the respective ``missing" matrix elements
 \begin{eqnarray}
 %\flushleft
  \Theta_{2,N}\,,\nonumber \\
  \Theta_{2,N-1}\,,\Theta_{3,N}\,,\nonumber \\
 \ldots\,,\nonumber \\
 \Theta_{2,2}\,,\Theta_{3,3}\,,\Theta_{4,4}\,,\ldots\,, \Theta_{N,N}\,
 \label{tiralbe}
 \end{eqnarray}
in recurrent manner.
\end{thm}
 \bp
Once we revealed the diagonal-wise-arranged recurrent pattern it is
easy and  entirely straightforward to verify its validity by the
corresponding trivial rearrangement of the two matrix
multiplications in Eq.~(\ref{dieudo}).
 \ep

\begin{pozn}
The diagonal-wise recurrent nature of Eq.~(\ref{dieudo}) given by
Theorem 1 has only been revealed by the {\it post factum}
inspection of the results of Ref.~\cite{gegenb}.
\end{pozn}

\section{The positive definiteness of the metric\label{duik}}

During the recent years we are witnessing the remarkable growth of
popularity of the building of quantum models which combine the
``false" non-Hermiticity $H \neq H^\dagger$ of the comparatively
elementary Hamiltonian acting in a ``friendly" Hilbert space ${\cal
H}^{(F)}$ with the simultaneous ``sophisticated" Hermiticity $H =
H^\ddagger$ of the same Hamiltonian in another,  less usual,
amended, ``standard" Hilbert space ${\cal H}^{(S)}$ endowed with a
nontrivial metric $\Theta= \Theta^{(S)}\neq I$. The essence of such
an innovation becomes entirely transparent and obvious when one
eliminates the partial confusion caused by the traditional
terminology. The key point is that one never leaves the abstract
theoretical framework of quantum theory. Just a few new mathematical
tricks (like, typically, an unusual, non-unitary generalization of
the most common Fourier transformation) are being added to the
traditional textbook recipes.

\subsection{The formulation of quantum theory using an {\em ad hoc} triplet of
complementary Hilbert spaces}

The Dieudonn\'e-equation constraint imposed on a Hamiltonian $H$ is
in fact equivalent to the manifest Hermiticity of its isospectral
image
 \be
 \mathfrak{h}=\Omega\,H\,\Omega^{-1}=\mathfrak{h}^\dagger\,
 \label{ouzo}
 \ee
\cite{SIGMA}. In principle (though not always in practice) the
latter operator is  defined as acting in the physical Hilbert space
${\cal H}^{(P)}$ in which the traditional, trivial metric
$\Theta^{(P)}=I$ is being used.

Both the Hilbert spaces ${\cal H}^{(P)}$ and ${\cal H}^{(S)}$ may be
perceived as unitary equivalent. We may deduce
 \be
 \mathfrak{h}^\dagger=\left (\Omega^{-1}
 \right )^\dagger\,H^\dagger\,\Omega^\dagger\,.
 \label{spouzo}
 \ee
After we abbreviate $ \Omega^\dagger \Omega:=\Theta$ we end up with
the Dieudonn\'e's  relation (\ref{dieudo}).

\subsection{The Gegenbauerian illustrative example}

For our real and finite-dimensional Gegenbauerian Hamiltonians
$H=H^{(N)}(a)$ which are given in advance, the Dieudonn\'e's
relation (\ref{dieudo}) forms the set of $N^2$ constraints imposed
upon the $[N(N+1)/2]-$plet of the unknown real matrix elements of
the metric matrix $\Theta=\Theta^\dagger>0$. In papers \cite{fund} -
\cite{posledni}  we proposed the non-numerical,
symbolic-manipulation  approach to constructions of a complete
solution of this linear algebraic system. What remains for us to
construct is the appropriate domain ${\cal D}$ of free parameters
for which these candidates for the metric remain positive definite,
i.e., truly eligible in the appropriate definitions of the
generalized Hermitian conjugation and/or of the appropriate
Hilbert-space inner product.

In Ref.~\cite{gegenb} we discussed a few specific examples of
candidates $\Theta^{(N)}(a)$ for the Gegenbauerian metrics. We
revealed that such a study leads  to a purely numerical description
supporting the hypothesis that the domains ${\cal D}^{(N)}(a)$
change  ``smoothly" with $N$ and lead to the non-empty and
sufficiently large limiting domains $\lim_{N \to \infty}{\cal
D}^{(N)}(a) ={\cal D}^{(\infty)}(a) \neq \emptyset$.

A comparatively weak $N-$dependence characterizes even the domains
${\cal D}^{(N)}(a)$ at small  $N \gtrapprox 5$. The direct
evaluation of the eigenvalues of $\Theta^{(N)}(a)$ (i.e., the more
precise determination of the boundaries  $\partial {\cal
D}^{(N)}(a)$) only suffers of the errors caused by the
multiple-scale nature of these eigenvalues.

In Ref.~\cite{gegenb} we were only able to provide a transparent
graphical illustration of the free-parameter-dependence of the
spectrum of selected $\Theta^{(N)}(a)$s at the very first nontrivial
dimension $N=3$. In the context of programming in MAPLE (offering an
adaptable floating-point precision arithmetics) the remedy is easy.
One may take, say, the $N=3$ toy metric of Ref.~\cite{gegenb},
 $$
 \Theta_g^{(3)}(a)=
 \left[ \begin {array}{ccc} 2\,{a}^{2}&2\,ga&0\\\noalign{\medskip}2\,g
a&a+1&g\\\noalign{\medskip}0&g&{\frac {a+2}{2\,a+1}}\end {array}
 \right]
 $$
and represent the triplet of eigenvalues $p_j(g)$, $j = 1,2,3$ (as
sampled in Figure Nr. 3 of {\it loc. cit.}  at $a=1$) in logarithmic
scale yielding, say, the adapted $a=1$ secular equation
 $$
 \det \, \left[ \begin {array}{ccc} 2-{e^{-{\it ttr}-20}}&2\,g&0\\
 \noalign{\medskip}2\,g&2-{e^{-{\it ttr}-20}}&g\\\noalign{\medskip}0&
g&1-{e^{-{\it ttr}-20}}\end {array} \right]=0\,
$$
i.e., the non-polynomial version of our eigenvalue problem,
 $$
 4-8\,{e^{-{\it ttr}-20}}+5\, \left( {e^{-{\it ttr}-20}} \right) ^{2}-6
\,{g}^{2}- \left( {e^{-{\it ttr}-20}} \right) ^{3}+5\,{g}^{2}{e^{-{
\it ttr}-20}}=0\,.
$$
In a test run the numerical analysis of this equation reproduced the
results given in Table Nr. 1 of {\it loc. cit.}.

\begin{figure}[h]                     %instead of \begin{figure}[t]
\begin{center}                         %instead of \begin{center}
\epsfig{file=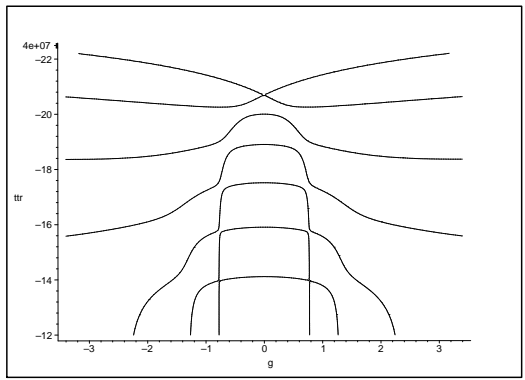,angle=270,width=0.5\textwidth}
\end{center}                         %instead of \end{center}
\vspace{-2mm}\caption{Seven eigenvalues $p=p(g)$ of metric
$\Theta^{(7)}_g(1)$.
 \label{fi1}}
\end{figure}

\begin{figure}[h]                     %instead of \begin{figure}[t]
\begin{center}                         %instead of \begin{center}
\epsfig{file=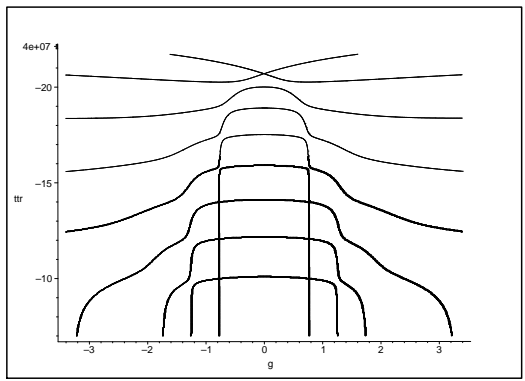,angle=270,width=0.5\textwidth}
\end{center}                         %instead of \end{center}
\vspace{-2mm}\caption{Nine eigenvalues $p=p(g)$ of metric
$\Theta^{(9)}_g(1)$.
 \label{fi2}}
\end{figure}

On this basis one may expect that the key problem brought by the
rescaling appears tractable by the MAPLE-based numerical software.
The main gain came with the substantial extension of the feasibility
of the graphical determinations of the parameter-dependence of the
physical domains ${\cal D}^{(N)}_g(a)$ with the growth of $N$. The
characteristic illustration is offered  by Figures \ref{fi1} and
\ref{fi2} which clearly demonstrate the emergence of an obvious
pattern which was not accessible without rescaling.

%********** Figure 1 zde

%********** Figure 2 zde

\section{Summary}

From the point of view of Quantum Mechanics it is rather unfortunate
that for a given Hamiltonian $H$ the specification of the metric
$\Theta(H)$ prescribed by Dieudonn\'e Eq.~(\ref{dieudo}) is
ambiguous \cite{Geyer}. In this context, our series of papers
\cite{fund} - \cite{posledni} has been devoted to the constructive
study of the one-to-many mappings $H \to \Theta(H)$. In essence, we
offered there a new methodical recipe of a systematic suppression of
the ambiguity of the menu of eligible $\Theta(H)$s.

In our present continuation and extension of this paper we decided
to explain the symbolic-manipulation aspects of such a recipe in
more detail. Emphasizing that such a problem would be hardly
tractable and/or solvable without an essential interaction between
the abstract quantum theory and the symbolic-manipulation techniques
and algebraic constructionis assisted by contemporary computers.

One of byproducts of such an interaction between methods has been
shown to lie in the possible amendments of the numerical aspects of
the necessary simultaneous analysis and study of  physical domains
and of the other properties of {\em both} of the
physics-representing operators $H$ and $\Theta$. On the basis of
these results one can conclude that the  confirmation of feasibility
of a methodical symbiosis between algebra and analysis seems to be
able to contribute to the contemporary quick growth of popularity of
the practical phenomenological applications of quantum models with
nontrivial metrics, e.g., in optics \cite{Makris}.

\section*{Appendix A. A compact review of Quantum Mechanics using
nontrivial metrics $\Theta \neq I$\label{secujedna}}

In the spirit of certain textbooks on Quantum Mechanics
\cite{Messiah} the Dirac's ket vectors  $|\psi\kt \in {\cal V}$ may
be interpreted as describing different quantum systems the nature of
which may vary with the other characteristics of the system in
question. Thus, one can start from the most common single-particle
(or, in some applications, single quasi-particle) quantum models
where, roughly speaking, there exists an operator $\hat{Q}$ of the
particle position with eigenvalues $q\in \mathbb{R}$ and eigenkets
$|q\kt$. One then introduces the standard (often called Dirac's)
Hermitian conjugation (i.e., transposition plus complex conjugation)
 \be
 {\cal T}^{(Dirac)}: |\psi\kt \ \to \ \br \psi|\,.
 \label{diraco}
 \ee
In this manner one represents the kets of states $|\psi\kt$ by the
concrete square-integrable functions $\psi(q) = \br q|\psi\kt$ (of
the coordinate $q$) called wave functions. In this context the
discrete, finite-dimensional models exemplified by Eq.~(\ref{input})
may be perceived as the most convenient testing ground of the
methods.

In all of the similar classes of examples the vector space of states
${\cal V}$ becomes endowed with the most common inner product
 \be
 \br \psi_a|\psi_b\kt = \int_{-\infty}^\infty\,\psi_a^*(q)\psi_b(q)\
 dq
 \label{calling}
 \ee
with, possibly, the integration replaced by the infinite or finite
summation. Thus, we may (and usually do) set ${\cal V}\equiv
L^2(\mathbb{R})$, etc.

Without any real danger of misunderstanding we may speak here about
the ``friendly" Hilbert space of states ${\cal H}^{(F)} \equiv {\cal
V}$, calling the variable $q$ in Eq.~(\ref{calling}) ``the
coordinate". In parallel we usually perform a maximally convenient
choice of the Hamiltonian $H$ based on the so called principle of
correspondence which ``dictates" us to split $H=T+V$ where the
general interaction operator $V$ is represented, say, by a kernel
$V(q,q')$ when acting upon the wave functions. In practice, this
kernel is most often chosen as proportional to the Dirac's
delta-function so that $V$ becomes an elementary multiplicative
operator $V =V_{local}= V(q)$. Similarly, the most popular and
preferred form of the ``kinetic energy" $T$ is a differential
operator, say, $T = T_{local}= -d^2/dq^2$ in single dimension and
suitable units.

A word of warning emerges when we perform a Fourier transformation
in ${\cal H}^{(F)} $ so that the variable $q$ becomes replaced by
$p$ (= momentum). One should rather denote the latter,
``Fourier-image" space by the slightly different symbol ${\cal
H}^{(P)} $, therefore (with the superscript still abbreviating
``physical" \cite{SIGMA}). Paradoxically, after this change of frame
the kinetic operator $T_{local}$ becomes multiplicative while
$V_{local}$ becomes strongly non-local in momenta.

The truly deep change of the paradigm comes with the models where
the necessity of the observability of the coordinate $q$ is
abandoned completely. One may still start from the vector space of
kets ${\cal V}$ and endow it with the Hilbert-space structure  via
the similarly looking inner product
 \be
 \br \psi_a|\psi_b\kt = \int_{\cal C}\,\psi_a^*(s)\psi_b(s)\
 ds\,.
 \label{recalling}
 \ee
The non-existence of the position operator $\hat{Q}$ changes the
physics of course. The key point is that we lose the one-to-one
correspondence between the integration path ${\cal C}$ and the
spectrum $\mathbb{R}$ of any coordinate-mimicking operator. The
physics-independent optional variable $s$ remains purely formal.

In such a setting our choice of the physical observables remains
unrestricted and is entirely arbitrary. It need not be related to
any usual classical system, either. For illustration one might
recall the pedagogically motivated paper~\cite{Hoo}. In a slightly
provocative demonstration of the abstract nature of quantum theory
the variable $s$ in Eq.~(\ref{recalling}) has been chosen there as
an observable ``time" of a hypothetical ``quantum clock" system.

Definition (\ref{recalling}) of the inner product in ${\cal
H}^{(F)}$ or in ${\cal H}^{(P)}$ is the theoretical framework within
which the traditional quantum mechanics works, reducing the full
power of the theory to something acceptable via analogies with
classical physics. Still, one need not move too far. Even in
Ref.~\cite{fund}, for example, we did not treat the variable $s$ in
Eq.~(\ref{recalling}) as a purely formal quantity, having preserved
at least a part of its relation to the position $q$. In particular,
we stayed in the middle of the path towards abstraction and worked
still with the usual one-dimensional ordinary differential
Schr\"{o}dinger equation
 \be
 -\frac{d^2}{ds^2} \,\psi_n (s) + V(s)
  \,\psi_n (s)= E_n \,\psi_n (s)\,,\ \ \ \ \ n = 0, 1, \ldots\,
   \label{SE}
 \ee
where the key nonstandard features can be seen in

\begin{itemize}

\item
the admissibility of the complex potentials sampled by the
power-law-anharmonic family $V(s) = -({\rm i} s)^{2+\delta}$ of
Ref.~\cite{BBjmp} giving the ``standard" real and discrete
bound-state spectra at any $\delta>0$ (cf. the proofs in
\cite{DDT});

\item
the admissibility of the replacement of the usual real line of $s$
by a suitable complex curve ${\cal C}=\mathbb{C}(\mathbb{R})$ which
may even be, in principle, living on a complicated multisheeted
Riemann surface \cite{tobog};

\item
in the possibility of a systematic study of its discrete analogues
and simplifications.

\end{itemize}

 \noindent
Naturally, one partially leaves the more or less safe guidance
offered by the principle of correspondence. Just a partial
revitalization of this guidance is possible (cf., e.g., a nice
example-based discussion of this point in Ref.~\cite{cubic}). A
partial reward for this loss can be sought in a new tractability of
some traditionally contradictory quantization problems (say, when
working, say, with the relativistic Klein-Gordon equations
\cite{KG}).

The main theoretical difficulty consists in the vast ambiguity of
the necessary appropriate generalization of Eq.~(\ref{diraco}). The
general recipe (explained already in \cite{Geyer} or, more
explicitly, in \cite{SIGMA}) is metric-dependent and reads
 \be
 {\cal T}^{(general)}_\Theta: |\psi\kt \ \to \ \bbr \psi|:=\br
 \psi|\Theta\,.
 \label{nediraco}
 \ee
This means that using the language of wave functions $\psi(s)$ with
$s \in {\cal C}$ we must replace the most common single-integral
definition (\ref{recalling}) of the inner product in the original
``friendly" Hilbert space ${\cal H}^{(F)}$ by the more sophisticated
double-integral formula
 \be
 \bbr \psi_a|\psi_b\kt = \int_{\cal C}\,\int_{\cal C}\,
 \psi_a^*(s)\,\Theta(s,s')\,\psi_b(s')\
 ds\,ds'\,.
 \label{birecalling}
 \ee
In terms of an integral-operator-kernel representation
$\Theta(s,s')$ of our abstract metric operator
$\Theta=\Theta^\dagger>0$ this recipe defines a more sophisticated
inner product which converts the ket-vector space ${\cal V}$ into
the metric-dependent (and physics-representing, ``standard"
\cite{SIGMA}) Hilbert space ${\cal H}^{(S)}$ .

%=========================

%\newpage

%\vspace{15mm}


\begin{thebibliography}{00}


\bibitem{fund}
M. Znojil,
%Fundamental length in quantum theories with PT-symmetric
%Hamiltonians.
Phys. Rev. D. 80, 045022 (2009).
%(13
%pages) (arXiv:0907.2677 [hep-th] 15 Jul 2009). (No. 4, 21 August 2009)


\bibitem{gegenb}
M. Znojil,
%Gegenbauer-solvable quantum chain model
Phys. Rev. A 82, 052113  (2010).
%http://dx.doi.org/10.1103/PhysRevA.82.052113
%(arXiv:1011.4803)




\bibitem{...}
%\bibitem{smeared}
%[DD2]
M. Znojil, %Scattering theory using smeared non-Hermitian
%potentials.
Phys. Rev. D. 80, 045009 (2009); %(No. 4, 12 August 2009)(12
%pages) (arXiv:0903.1007v2 [quant-ph] 29 Jun 2009).

M. Znojil,
%[DD1]  Cryptohermitian picture of scattering using
%quasilocal metric operators.
SYMMETRY, INTEGRABILITY and GEOMETRY: METHODS and APPLICATIONS
%SIGMA
5, 085 (2009). %, 21 pages
%(arXiv:0908.4045 [quant-ph] 27 Aug 2009)



\bibitem{determ}
M. Znojil,
%Determination of the domain of the admissible matrix elements in the four-dimensional PT-symmetric anharmonic model
Phys. Lett. A 367, 300 (2007).
%, Issues 4-5, 30 July 2007, Pages 300-306 (available online)
%(quant-ph/0703168).





\bibitem{fundgra}
M. Znojil, %Fundamental length in quantum theories with PT-symmetric
%Hamiltonians II: The case of quantum graphs.
Phys. Rev. D. 80, 105004 (2009);
%(No. 10, 15 November 2009, 20 pages)
%http://dx.doi.org/10.1103/PhysRevD.80.105004 (arXiv:0910.2560
%[hep-th] 14 Oct 2009)
%
%
%Pseudo-Hermitian continuous-time quantum walks

S. Salimi and A. Sorouri,
%Journal of Physics A Mathematical and Theoretical 43, 275304 (2010)
J. Phys. A: Math. Gen. 43, 275304 (2010).


\bibitem{anomal}
M. Znojil,
J. Phys. A: Math. Theor. 43, 335303 (2010).





\bibitem{posledni}
M. Znojil, %Complete set of inner products for a discrete
%PT-symmetric square-well Hamiltonian.
J. Math. Phys. 50, 122105 (2009).
%, http://dx.doi.org/10.1063/1.3272002, (arXiv:0911.0336v1
%[math-ph] 2 Nov 2009)

\bibitem{Dieudonne}
 J. Dieudonne, %Quasi-Hermitian operators,
  Proc. Int. Symp.
            Lin. Spaces (Pergamon, Oxford, 1961), p. 115.
            %-122


\bibitem{Geyer}
F. G. Scholtz, H. B. Geyer and F. J. W. Hahne, Ann. Phys. (NY) 213,
 74 (1992).
%% Quasi-Hermitian Operators in Quantum Mechanics
%% and the Variational Principle,


\bibitem{Carl}
%Bender C M 2007 Making sense of non-hermitian Hamiltonians Rep.
%Prog. Phys. 70 947-1018
C. M. Bender, Rep. Prog. Phys. 70, 947 (2007).
%, submitted (hep-th/0703096).

\bibitem{SIGMA}
M. Znojil,
%Three-Hilbert-space formulation of Quantum Mechanics
SYMMETRY, INTEGRABILITY and GEOMETRY: METHODS and APPLICATIONS %SIGMA
5,  001 (2009).
%, 19 pages;


\bibitem{Maple}
B. W. Char et al, Maple V Language Reference Manual (Springer, New
York, 1993).

\bibitem{Ryzhik}
I. S. Gradshteyn and I. M. Ryzhik, Tablicy integralov, summ, ryadov
i proizvedenii (Nauka, Moscow, 1971).

\bibitem{Stegun}
M. Abramowitz and I. A. Stegun, Handbook of Mathematical Functions
(Dover, New York, 1970).



\bibitem{Williams}
J. P. Williams,
% Operators Similar to their Adjoints,
Proc. Amer.
            Math. Soc. 20, 121 (1969);
            %-123


%\bibitem{sie}
P. Siegl,  J. Phys. A: Math. Theor. 41, 244025 (2008).


\bibitem{Dorey}
P. Dorey, C. Dunning  and R. Tateo,  %The ODE/IM correspondence,
J. Phys. A: Math. Theor. 40,  R205 (2007);
%-R283, hep-th/0703066;

E. B. Davies,  Linear operators and their spectra (Cambridge,
Cambridge University Press, 2007).

\bibitem{ali}
A. Mostafazadeh, Pseudo-Hermitian Quantum Mechanics,
%
arXiv:0810.5643,
%.
% arXiv:0810.5643 [pdf, ps, other]
%Title: Pseudo-Hermitian Representation of Quantum Mechanics Authors:
%Ali Mostafazadeh Comments: 76 pages, 2 figures, 243 references,
%revised version
to appear in Int. J. Geom. Meth. Mod. Phys. % Subjects:
%Quantum Physics (quant-ph); High Energy Physics - Theory (hep-th);
%Mathematical Physics (math-ph)

\bibitem{Makris}
%Beam Dynamics in PT Symmetric Optical Lattices
%
K. G. Makris, R. El-Ganainy, D. N. Christodoulides and Z. H.
Musslimani, Phys. Rev. Lett. 100, 103904 (2008);


%PT-symmetric optical lattices
%
%Konstantinos
K. G. Makris, R. El-Ganainy, D. N. Christodoulides and Z. H.
Musslimani,
%
Phys. Rev. A 81, 063807 (2010);

%Phys. Rev. A 82, 010103(R) (2010) [4 pages] PT optical lattices and
%universality in beam dynamics AbstractReferences No Citing Articles
%Download: PDF (491 kB) Export: BibTeX or EndNote (RIS) Mei C.
%Zheng1, Demetrios N. Christodoulides2, Ragnar Fleischmann3, and
%Tsampikos Kottos1
M. C. Zheng, D. N. Christodoulides, R. Fleischmann and T. Kottos,
Phys. Rev. A 82, 010103 (2010);

%Observation of PT-Symmetry Breaking in Complex Optical Potentials
%
%
A. Guo, G. J. Salamo, D. Duchesne, R. Morandotti, M. Volatier-Ravat,
V. Aimez, G. A. Siviloglou and D. N. Christodoulides, Phys. Rev.
Lett. 103, 093902 (2009);

%Optical structures with local {\cal PT} -symmetry
O. Bendix, R. Fleischmann, T. Kottos and B. Shapiro, % Journal of
%Physics A Mathematical
%and Theoretical
J. Phys. A: Math. Theor. 43, 265305 (2010);


%Observation of parity–time symmetry in optics Christian
%
Ch. E. R\"{u}ter,
%Konstantinos
K. G. Makris, R. El-Ganainy, D. N. Christodoulides, M. Segev and D.
Kip, Nature Phys. 6, 192 (2010).

\bibitem{Messiah}
A. Messiah, Quantum Mechanics (North Holland, Amsterdam, 1961).



\bibitem{Hoo}
J. Hilgevoord, Am. J. Phys. 70, 301 (2002).

\bibitem{BBjmp}
C. M. Bender, S. Boettcher and P. M. Meisinger, J. Math. Phys. 40,
2201 (1999).

\bibitem{DDT}
P. Dorey, C. Dunning and R. Tateo, J. Phys. {A}: Math. Gen. 34, 5679
(2001).


\bibitem{tobog}
M. Znojil,
%Planarizable supersymmetric quantum toboggans.
SIGMA 7, 018  (2011).
%, 24 pages,
%doi:10.3842/SIGMA.2011.018
%arXiv:1102.5162


\bibitem{cubic}
A. Mostafazadeh, J. Phys. A: Math. Gen. 39, 10171 (2006).


\bibitem{KG}
M. Znojil,
%Relativistic supersymmetric quantum mechanics based on Klein-Gordon equation.
J. Phys. A: Math. Gen. 37, 9557 (2004).
%-9571
%(hep-th/0408232)



\end{thebibliography}
\end{document}